\documentclass[12pt]{iopart}
\jl{1}
\eqnobysec

\input amssym.def
\input amssym.tex

\usepackage{subeqn}

\newcommand{\vecvar}[1]{\mbox{\boldmath$#1$}}

\newcommand{\PTP}{{\em Prog. Theor. Phys.} }

\newcommand{\SAM}{{\em Stud. Appl. Math.} }

\newcommand{\CMP}{{\em Commun. Math. Phys.} }

\def\hf{\frac{1}{2}}
\def\beq{\begin{equation}} \def\eeq{\end{equation}}
\def\bseq{\begin{subequations}} \def\eseq{\end{subequations}}
\def\bea{\begin{eqnarray}} \def\eea{\end{eqnarray}}
\def\bsea{\begin{subeqnarray}} \def\esea{\end{subeqnarray}}

\let\ti=\tilde

\def\eql{\eqalign}

\let\nn=\nonumber
\def\beann{\begin{eqnarray*}} \def\eeann{\end{eqnarray*}}

\let\a=\alpha \let\be=\beta  \let\de=\delta
 \let\z=\zeta

\def\0{\over } \def\1{\vec }     \def\2{{1\over2}} \def\4{{1\over4}}
\def\5{\bar }  \def\6{\partial } \def\7#1{{#1}\llap{/}}

\def\<{\langle } \def\>{\rangle }

\def\i{{\rm i}}

\def\d{{\rm d}}

\def\e{{\rm e}}

\def\Q1{{\sf Q}}
\def\QS1{{\sf q}}
\def\qa{Q}
\def\qsa{q}

\def\R1{{\sf R}}
\def\RS1{{\sf r}}
\def\ra{R}
\def\rsa{r}

\def\P1{{\sf P}}
\def\pa{P}

\def\S1{{\sf S}}
\def\sa{S}

\def\ja{{\cal P}}

\def\ta{{\cal S}}

\begin{document}
\jl{5}

\title{Complete integrability of derivative 
nonlinear Schr\"{o}dinger-type equations}

\author{Takayuki Tsuchida\footnote[1]{E-mail address: 
{\tt tsuchida@monet.phys.s.u-tokyo.ac.jp}} 
and Miki Wadati 
%\footnote[3]{e-mail: {\tt wadati@monet.phys.s.u-tokyo.ac.jp}}
}

\address{Department of Physics, Graduate School of Science, \\
University of Tokyo, \\
Hongo 7--3--1, Bunkyo-ku, Tokyo 113--0033, Japan}

\begin{abstract}
We study matrix generalizations of derivative nonlinear 
Schr\"{o}dinger-type equations, which were shown by 
Olver and Sokolov to possess a
higher symmetry. We prove that two of them are
`C-integrable' and the rest of them are `S-integrable' in 
Calogero's terminology.
\end{abstract}

\pacs{02.10Jf, 02.30.Jr, 03.65.Ge, 11.30.-j}
\submitted

\maketitle

\section{Introduction}
It has been known that a large part of soliton equations in $1+1$
dimensions have integrable matrix generalizations, or more generally, 
integrable multi-field generalizations. Such generalizations have been
studied for a wide variety of single(or two)-component integrable
systems by virtue of various approaches such as Jordan algebras and 
Jordan pairs \cite{Svi1,Svi2,Svi3}, inverse scattering 
method \cite{Wadati3,Fordy2,Tsuchida1,Tsuchida3,Tsuchida5}, 
Hirota's method \cite{Iwao,Hirota2,Ohta1,Ohta2,Ohta3}. However, 
as far as the authors know, matrix generalizations of the 
derivative nonlinear Schr\"{o}dinger(DNLS)-type systems \cite{KN,CLL}
have not been studied throughly. 

In recent papers \cite{Olver1,Olver2}, Olver and Sokolov made a detailed 
investigation on the DNLS-type systems of the form
\beq
\eql{
P_t = P_{xx} + f(P,S,P_x,S_x)
\\
S_t = -S_{xx} + g(P,S,P_x,S_x). 
}
\label{DNLStype}
\eeq
Here $P$ and $S$ take values in an associative algebra. For simplicity,
in the following we regard $P$ and $S$ as matrix-valued. $f$ and $g$
are non-commutative polynomials of weight $5$, where the weights of 
$\6_t$, $\6_x$, $P$ and $S$ are respectively assigned to be 
$4$, $2$, $1$ and $1$. 
They made a complete list of the DNLS-type systems \eref{DNLStype} 
which have one higher symmetry of the following form with 
weight $9$,
\[
\eql{
P_\tau = P_{xxxx} + \ti{f}(P,S,P_x,S_x,P_{xx},S_{xx},P_{xxx},S_{xxx})
\\
S_\tau = -S_{xxxx} + \ti{g}(P,S,P_x,S_x,P_{xx},S_{xx},P_{xxx},S_{xxx}). 
}
\]
Here the commutativity of the two flows, i.e. $\6_t \6_\tau P = \6_\tau
\6_t P$, $\6_t \6_\tau S = \6_\tau \6_t S$, works as a strong 
constraint on the form of $f$ and $g$ in order for the non-commutative
polynomials $\ti{f}$ and $\ti{g}$ to exist. See 
ref. \cite{Olver3,Mikhailov,Fokas} for a more 
precise explanation of symmetry approach.
% in terms of the Fr\'echet derivative. 

The entries of the list are divided into two kinds. Two systems in the
list are interpreted as non-abelian analogues of the following 
integrable system
\beq
\eql{
p_t = p_{xx} + 2\a p^2 s_x + 2 \a p p_x s -\a \be p^3 s^2
\\
s_t = -s_{xx} + 2\be s^2 p_x + 2\be ss_x p+ \a \be s^3 p^2 
}
\label{Csys}
\eeq
for a particular choice of the constants $\a$ and $\be$. The system 
\eref{Csys} is linearizable by a change of the dependent
variables. Thus, we can construct the general solution of the 
system. We often refer to such linearizable systems as `C-integrable'
in Calogero's terminology \cite{Calo1}. On the other hand, excepting
 the two entries corresponding to \eref{Csys}, the scalar-valued 
counterparts of the systems in the list by Olver and Sokolov are 
given by \cite{Tsuchida5}
\beq
\eql{
 \i q_t + q_{xx} +4\i \de q^2 r_x + \i (4\de -\a) qq_x r
+ \de (4\de + \a) q^3 r^2 =0
\\
\i r_t - r_{xx}  +4\i \de r^2 q_x + \i (4\de-\a) rr_x q
- \de (4\de+\a) r^3 q^2  =0
}
\label{K_eq}
\eeq
for special choices of the constants $\a$ and $\de$. This system was 
generated via a gauge transformation for the DNLS equation 
%(\eref{} with $\a= , \de=$) 
by Kund \cite{Kund}. We can write down a Lax pair for the system
\eref{K_eq} with the help of the gauge transformation. According to
Calogero's terminology, this kind of systems, which are
linearizable by the inverse scattering formulation, are called 
`S-integrable' systems \cite{Calo1}. 

If a system has one higher symmetry, it has been believed that the 
system has an infinite series of symmetries and thus is completely 
integrable. However, a system proposed by Bakirov was recently proved
to be a counter-example to this empirical law \cite{Beukers}. Thus,
there is no guarantee that the systems in the list by 
Olver and Sokolov are really integrable, although the counter-example
seems very exceptional. 

The aim of this paper is to establish the complete integrability of
all the matrix-valued systems given in \cite{Olver2}. In a 
previous paper \cite{Tsuchida5}, the authors introduced a Lax pair for 
the matrix generalization of the Chen-Lee-Liu equation \cite{CLL},
\beq
\eql{
 \i \Q1_t + \Q1_{xx} -\i \Q1 \R1 \Q1_x = O
\\
\i \R1_t - \R1_{xx} -\i \R1_x \Q1 \R1 = O
}
\label{ori}
\eeq
which is a member of the list by Olver and Sokolov. In the present
paper, we generalize the Lax pair for \eref{ori} to be applicable for
several matrix systems of the DNLS type in \cite{Olver2}. 

The paper consists of the following. In section 2, we extend a 
notion of the gauge transformation to the non-abelian case. We give an 
explicit expression of the Lax pairs for all systems but two in
\cite{Olver2} to show their `S-integrability'. On the other hand, in
section 3, we prove the remaining two systems to be linearizable 
so that they are `C-integrable'. Concluding remarks are
given in the last section, section 4. Throughout this paper, capital 
letters denote matrices while small letters represent scalars. 
Equation numbers without `section-number$.$' refer to equations 
in \cite{Olver2}.

\section{`S-integrable' systems}
\label{S-int}

We begin with a brief summary of the matrix generalization 
of the Chen-Lee-Liu-type DNLS equation \cite{Tsuchida5}. 
A system of linear differential equations for a 
vector $\Psi$
\beq
\Psi_x = U \Psi \hspace{6mm} \Psi_t = V \Psi
\label{scattering}
\eeq
is consistent if the following equation is satisfied:
\beq
U_t -V_x +UV-VU = O. 
\label{Lax_eq}
\eeq
The matrices $U$ and $V$ are called Lax matrices or a Lax pair. 
We introduce the following form of the Lax pair,
\bea
\fl U =
\i \z^2 \left[
\begin{array}{cc}
 -I_1  &  \\
    &  I_2 \\
\end{array}
\right]
+ 
\z \left[
\begin{array}{cc}
   &  \Q1 \\
 \R1  &   \\
\end{array}
\right]
+ \i
\left[
\begin{array}{cc}
  O &  \\
    & \frac{1}{2}\R1 \Q1  \\
\end{array}
\right]
\label{U_form}
%\\
%\vspace{-10mm}
%\nn 
\\
\fl V =
\i \z^4 
\left[
\begin{array}{cc}
 -2I_1 &  \\
   & 2I_2  \\
\end{array}
\right]
+\z^3
\left[
\begin{array}{cc}
  & 2\Q1 \\
 2\R1 &  \\
\end{array}
\right]
+\i \z^2
\left[
\begin{array}{cc}
-\Q1 \R1  &  \\
  & \R1 \Q1 \\
\end{array}
\right]
\nn \\
\fl  \hspace{7.5mm} +\z
\left[
\begin{array}{cc}
  & \i \Q1_x + \frac{1}{2} \Q1 \R1 \Q1 \\
 -\i \R1_x + \frac{1}{2}\R1 \Q1 \R1 &  \\
\end{array}
\right]
+\i
\left[
\begin{array}{cc}
 O &  \\
  & \frac{\i}{2}(\R1 \Q1_x - \R1_x \Q1) + \frac{1}{4}\R1 \Q1 \R1 \Q1 \\
\end{array}
\right] .
% \nn \\
% &&
\label{V_form}
\eea
Here, $\z$ is the spectral parameter. $I_1$ and $I_2$ are respectively 
the $n_1 \times n_1$ and the $n_2 \times n_2$ identity matrices. $\Q1$ is an 
$n_1 \times n_2$ matrix and $\R1$ is an $n_2 \times n_1$ matrix. Putting
\eref{U_form} and \eref{V_form} into \eref{Lax_eq}, 
we obtain the matrix version of the Chen-Lee-Liu-type DNLS equation
\beq
\eql{
 \i \Q1_t + \Q1_{xx} -\i \Q1 \R1 \Q1_x = O
\\
\i \R1_t - \R1_{xx} -\i \R1_x \Q1 \R1 = O.
}
\label{matrix_eq}
\eeq
It is to be noted that there is no restriction on the sizes of $\Q1$
and $\R1$, that is, on $n_1$ and $n_2$. The system \eref{matrix_eq} was 
shown to possess at least one higher symmetry
 \cite{Olver1,Olver2}. The system is now proved to be completely
 integrable in the sense that it has a Lax pair and, as a result, an 
infinite number of conservation laws.

Next, we shall prove the complete integrability of other systems in 
\cite{Olver2} in the same sense. For this purpose, we introduce a 
transformation of dependent variables:
\beq
\Q1 = F^{-1} \qa G^{-1} \hspace{5mm} \R1 = G \ra F
\label{tra1}
\eeq
or equivalently
\[
\qa = F\Q1 G \hspace{5mm} \ra = G^{-1} \R1 F^{-1}.
%\label{tra2}
\]
Here $F$ and $G$ are invertible matrices, which in general depend 
on $\qa$ and $\ra$ (or $\Q1$ and $\R1$). Then, time evolution
equations for $\Q1$ and $\R1$, \eref{matrix_eq}, are cast into 
those for $\qa$ and $\ra$:
\bseq
\bea
\fl \i \qa_t + \qa_{xx} - \i \qa\ra\qa_x 
-\i F_t F^{-1}\qa + \i \qa (G^{-1})_t G
- (F_x F^{-1})_x \qa + \qa \{ (G^{-1})_x G \}_x 
\nn \\
\fl 
-2 F_x F^{-1} \qa_x
+ 2\qa_x (G^{-1})_x G -2 F_x F^{-1}\qa (G^{-1})_x G + (F_x F^{-1})^2 \qa
\nn \\ 
\fl
+ \qa\{(G^{-1})_x G\}^2 + \i \qa\ra F_x F^{-1} \qa 
-\i \qa\ra\qa (G^{-1})_x G = O
\label{}
\eea
\bea
\fl \i \ra_t -\ra_{xx} -\i \ra_x \qa\ra -\i (G^{-1})_t G\ra + \i \ra F_t F^{-1}
+ \{ (G^{-1})_x G\}_x \ra -\ra (F_x F^{-1})_x 
\nn \\
\fl
+ 2(G^{-1})_x G \ra_x 
- 2\ra_x F_x F^{-1} + 2 (G^{-1})_x G \ra F_x F^{-1} - \{ (G^{-1})_x G \}^2 \ra 
\nn \\
\fl
- \ra(F_x F^{-1})^2 + \i (G^{-1})_x G \ra\qa\ra 
-\i \ra F_x F^{-1} \qa\ra = O .
\label{}
\eea
\label{trans_eq}
\eseq
A sufficient condition for \eref{trans_eq} to be 
local and closed equations is that $F_x F^{-1}$, $F_t F^{-1}$, 
$(G^{-1})_x G$ and $(G^{-1})_t G$ are expressed locally in closed 
forms in terms of $\qa$ and $\ra$, i.e. they do not include terms with 
integrals, infinite sums, etc. We impose this condition on $F$ and
$G$ in what follows. A closed expression of the Lax pair for the 
transformed system is given by performing the gauge transformation
\[
\Psi = g \Phi \hspace{5mm} \; 
g = \left[
\begin{array}{cc}
 F^{-1}  &  \\
  & G \\
\end{array}
\right].
\]
Due to this transformation, the linear problem 
%\eref{scattering} 
and the Lax pair for \eref{matrix_eq} are 
changed into those for \eref{trans_eq}:
%which changes the Lax matrices $U$, $V$ into
%
\[
\Phi_x = U' \Phi \hspace{6mm} \Phi_t = V' \Phi
\]
\bea
\fl U' &=& g^{-1} U g - g^{-1} g_x 
\nn \\
\fl &=&
\i \z^2 \left[
\begin{array}{cc}
 -I_1  &  \\
   &  I_2 \\
\end{array}
\right]
+ 
\z \left[
\begin{array}{cc}
   &  \qa \\
 \ra  &   \\
\end{array}
\right]
+ \i
\left[
\begin{array}{cc}
 -\i F_x F^{-1}  &  \\
  & \frac{1}{2}\ra\qa -\i (G^{-1})_x G \\
\end{array}
\right]
\label{}
\eea
\bea
\fl V' &=& g^{-1} V g - g^{-1} g_t 
\nn \\
\fl &=&
\i \z^4 
\left[
\begin{array}{cc}
 -2I_1 &  \\
   & 2I_2  \\
\end{array}
\right]
+\z^3
\left[
\begin{array}{cc}
  & 2\qa \\
 2\ra &  \\
\end{array}
\right]
+\i \z^2
\left[
\begin{array}{cc}
-\qa\ra  &  \\
  & \ra\qa \\
\end{array}
\right]
%\hspace{7.5mm} 
+\z
\left[
\begin{array}{cc}
  & V_{12} \\
 V_{21} & \\
\end{array}
\right]
\nn \\
\fl && 
+\i
\left[
\begin{array}{cc}
 -\i F_t F^{-1} &  \\
  & V_{22} \\
\end{array}
\right]  .
\label{}
\eea
Here 
\bea
\eql{
V_{12} = \i \qa_x + \hf \qa\ra\qa -\i F_x F^{-1} \qa + \i \qa (G^{-1})_x G 
\\
V_{21} = -\i \ra_x + \hf \ra\qa\ra + \i (G^{-1})_x G \ra -\i \ra F_x F^{-1} 
\\
V_{22} = \i \frac{1}{2}(\ra \qa_x - \ra_x \qa) + \frac{1}{4}\ra\qa\ra\qa 
-\i \ra F_x F^{-1} \qa 
 + \i \hf \ra\qa (G^{-1})_x G 
\\
\hspace{11mm}  + \i \hf (G^{-1})_x G \ra\qa -\i (G^{-1})_t G .
}
\nn
\eea
The above transformation is a powerful tool; it yields new integrable 
systems of the DNLS type by appropriate choices of $F$ and $G$. 
To confirm this, we list six illustrative examples (a)--(f) with 
the definition of $F$ and $G$, the evolution equations for $\qa$ and 
$\ra$ and the transformed Lax matrix $U'$:
\[
\fl {\rm (a)}\hspace{20mm} 
F = I_1
\]
\[
(G^{-1})_x = -\i \hf G^{-1} \R1 \Q1 = -\i \hf \ra\qa G^{-1}
\]
\bea
(G^{-1})_t &=& G^{-1} \Bigl\{ \hf (\R1 \Q1_x -\R1_x \Q1) 
-\i \frac{1}{4}\R1\Q1\R1\Q1 
\Bigr\} 
\nn \\
&=& \Bigl\{ \hf (\ra\qa_x - \ra_x \qa) -\i \frac{3}{4} \ra\qa\ra\qa 
  \Bigr\} G^{-1}
\nn
\eea
\beq
\eql{
\i \qa_t + \qa_{xx}-\i (\qa\ra\qa)_x = O
\\
\i \ra_t - \ra_{xx}-\i (\ra\qa\ra)_x = O
}
\label{a_eq}
\eeq
\beq
U' =
\i \z^2 \left[
\begin{array}{cc}
 -I_1  &  \\
   &  I_2 \\
\end{array}
\right]
+ 
\z \left[
\begin{array}{cc}
   &  \qa \\
 \ra  &   \\
\end{array}
\right].
\label{}
\eeq

\[
\fl {\rm (b)}\hspace{20mm} 
G^{-1} = I_2
\]
\[
F_x = -\i \hf F \Q1 \R1 = -\i \hf \qa\ra F
\]
\bea
F_t &=& F \Bigl\{ \hf (\Q1_x \R1 -\Q1\R1_x) -\i \frac{1}{4}
\Q1\R1\Q1\R1 \Bigr\}
\nn \\
&=& \Bigl\{ \hf (\qa_x \ra-\qa\ra_x) + \i \frac{1}{4} \qa\ra\qa\ra \Bigr\} F
\nn
\eea
\beq
\eql{
\i \qa_t + \qa_{xx} + \i \qa\ra_x \qa + \hf \qa\ra\qa\ra\qa =O
\\
\i \ra_t - \ra_{xx} + \i \ra\qa_x \ra -\hf \ra\qa\ra\qa\ra =O
}
\label{b_eq}
\eeq
\beq
U' =
\i \z^2 \left[
\begin{array}{cc}
 -I_1  &  \\
   &  I_2 \\
\end{array}
\right]
+ 
\z \left[
\begin{array}{cc}
   &  \qa \\
 \ra  &   \\
\end{array}
\right]
+ \i
\left[
\begin{array}{cc}
 -\frac{1}{2}\qa\ra  &  \\
    & \frac{1}{2}\ra\qa  \\
\end{array}
\right].
\label{}
\eeq

\[
\fl {\rm (c)}\hspace{20mm} 
F = I 
\]
\[
(G^{-1})_x = \i \hf \Q1\R1 G^{-1} = \i \hf \qa\ra G^{-1}
\]
\bea
(G^{-1})_t &=& \Bigl\{ \hf (\Q1\R1_x-\Q1_x \R1) + 
\i \frac{1}{4}\Q1\R1\Q1\R1 \Bigr\}G^{-1}
\nn \\
&=& \Bigl\{ \hf (\qa\ra_x-\qa_x \ra) + \i \frac{1}{4}\qa\ra\qa\ra 
     -\i \hf \qa^2 \ra^2 \Bigr\}G^{-1}
\nn
\eea
\beq
\eql{
\fl \i \qa_t + \qa_{xx} - \i \qa\ra\qa_x + \i \qa^2 \ra_x + \i \qa_x
\qa\ra -\hf \qa^2 \ra\qa\ra 
 + \hf \qa^3 \ra^2 + \hf \qa\ra\qa^2 \ra = O
\\
\fl \i \ra_t - \ra_{xx} -\i \ra_x \qa\ra + \i \qa_x \ra^2 + \i \qa\ra \ra_x + \hf \qa\ra \qa\ra^2 
-\hf \qa^2 \ra^3 - \hf \qa\ra^2 \qa\ra =O
}
\label{c_eq}
\eeq
\beq
U' =
\i \z^2 \left[
\begin{array}{cc}
 -I  &  \\
   &  I \\
\end{array}
\right]
+ 
\z \left[
\begin{array}{cc}
   &  \qa \\
 \ra  &   \\
\end{array}
\right]
+ \i
\left[
\begin{array}{cc}
  O &  \\
    & \frac{1}{2}(\ra\qa+\qa\ra)  \\
\end{array}
\right].
\label{}
\eeq

\[
\fl {\rm (d)}\hspace{20mm} 
G^{-1} = I
\]
\[
F_x = \i \hf \R1\Q1 F = \i \hf \ra\qa F
\]
\bea
F_t &=& \Bigl\{ \hf (\R1_x \Q1- \R1\Q1_x ) 
+ \i \frac{1}{4}\R1\Q1\R1\Q1 \Bigr\} F
\nn \\
&=& \Bigl\{ \hf (\ra_x \qa - \ra\qa_x) 
+ \i \frac{1}{4}\ra\qa\ra\qa + \i \hf \ra^2 \qa^2 
\Bigr\} F
\nn
\eea
\beq
\eql{
\i \qa_t + \qa_{xx} -\i \qa\ra\qa_x -\i \ra_x \qa^2 -\i \ra\qa\qa_x + \hf \ra^2 \qa^3
-\hf \qa\ra^2 \qa^2 =O
\\
\i \ra_t - \ra_{xx} -\i \ra_x \qa\ra -\i \ra^2 \qa_x -\i \ra_x \ra\qa -\hf \ra^3 \qa^2
+ \hf \ra^2 \qa^2 \ra =O
}
\label{d_eq}
\eeq
\beq
U' =
\i \z^2 \left[
\begin{array}{cc}
 -I  &  \\
   &  I \\
\end{array}
\right]
+ 
\z \left[
\begin{array}{cc}
   &  \qa \\
 \ra  &   \\
\end{array}
\right]
+ \i
\left[
\begin{array}{cc}
 \frac{1}{2}\ra\qa  &  \\
    & \frac{1}{2}\ra\qa  \\
\end{array}
\right].
\label{}
\eeq

\[
\fl {\rm (e)}\hspace{20mm} 
G^{-1} = F
\]
\[
F_x = -\i \hf F\R1\Q1 = -\i \hf \ra\qa F
\]
\bea
F_t &=& F \Bigl\{ \hf (\R1\Q1_x -\R1_x \Q1)
-\i \frac{1}{4}\R1\Q1\R1\Q1 \Bigr\}
\nn \\
&=& \Bigl\{ \hf (\ra\qa_x -\ra_x \qa) 
-\i \frac{3}{4} \ra\qa\ra\qa + \i \hf \ra^2
\qa^2 \Bigr\} F
\nn
\eea
\beq
\eql{
\fl \i \qa_t + \qa_{xx} -\i \qa\ra\qa_x 
-\i \qa\ra_x \qa 
-\i \qa_x \ra\qa + \i \ra_x \qa^2 + \i \ra\qa\qa_x -\ra\qa\ra\qa^2 
\\
\hspace{50mm} + \hf \ra^2 \qa^3 + \hf \ra\qa^2 \ra\qa =O
\\
\fl \i \ra_t - \ra_{xx} -\i \ra_x \qa\ra - \i \ra\qa_x \ra 
- \i \ra\qa\ra_x + \i \ra^2 \qa_x 
+ \i \ra_x \ra\qa + \ra^2 \qa\ra\qa 
\\
\hspace{50mm} - \hf \ra^3 \qa^2 -\hf \ra\qa\ra^2 \qa =O
}
\label{e_eq}
\eeq
\beq
U' =
\i \z^2 \left[
\begin{array}{cc}
 -I  &  \\
   &  I \\
\end{array}
\right]
+ 
\z \left[
\begin{array}{cc}
   &  \qa \\
 \ra  &   \\
\end{array}
\right]
+ \i
\left[
\begin{array}{cc}
 -\hf \ra\qa  &  \\
    & O  \\
\end{array}
\right].
\label{}
\eeq

\[
\fl {\rm (f)}\hspace{20mm} 
G^{-1} = F
\]
\[
F_x = -\i \hf F \Q1\R1 = -\i \hf \qa\ra F
\]
\bea
F_t &=& F \Bigl\{ \hf (\Q1_x \R1-\Q1\R1_x) -\i 
\frac{1}{4}\Q1\R1\Q1\R1 \Bigr\}
\nn \\
&=& \Bigl\{ \hf (\qa_x\ra-\qa\ra_x) + \i \frac{1}{4} 
\qa\ra\qa\ra -\i \hf \qa^2 \ra^2
\Bigr\} F
\nn
\eea
\beq
\eql{
\fl \i \qa_t + \qa_{xx} + \i \qa\ra_x \qa - \i \qa^2 \ra_x -\i \qa_x \qa\ra -\hf \qa^2 \ra^2 \qa
-\hf \qa^2\ra\qa\ra 
\\
\hspace{50mm}
+ \hf \qa^3 \ra^2 + \hf \qa\ra\qa\ra\qa =O
\\
\fl \i \ra_t -\ra_{xx} + \i \ra \qa_x \ra -\i \qa_x \ra^2 -\i \qa\ra\ra_x 
+ \hf \ra\qa^2 \ra^2 + \hf \qa\ra\qa\ra^2
\\
\hspace{50mm}
-\hf \qa^2 \ra^3 -\hf \ra\qa\ra\qa\ra =O
}
\label{f_eq}
\eeq
\beq
U' =
\i \z^2 \left[
\begin{array}{cc}
 -I  &  \\
   &  I \\
\end{array}
\right]
+ 
\z \left[
\begin{array}{cc}
   &  \qa \\
 \ra  &   \\
\end{array}
\right]
+ \i
\left[
\begin{array}{cc}
 -\frac{1}{2}\qa\ra  &  \\
    & \hf (\ra\qa - \qa\ra)  \\
\end{array}
\right].
\label{}
\eeq
For all of the six examples (a)--(f), the compatibility conditions for 
$F$ and $G^{-1}$, i.e.
\[
(F_x)_t = (F_t)_x \hspace{5mm} \{ (G^{-1})_x \}_t = \{ (G^{-1})_t \}_x
\]
can be checked by a straightforward calculation with the help of 
\eref{matrix_eq}. As is clear from the construction of the 
gauge transformations, $\qa$ and $\ra$ can be rectangular matrices 
for (a) and (b), while $\qa$ and $\ra$ must be square
matrices for (c)--(f). 

Comparing the above results with those by Olver and Sokolov, we find
that \eref{matrix_eq} and (10) in \cite{Olver2}, \eref{a_eq} and (7)
in \cite{Olver2}, \eref{b_eq} and (12) in \cite{Olver2}, \eref{c_eq} 
and (14) in \cite{Olver2}, \eref{d_eq} and (16) in \cite{Olver2}, 
\eref{e_eq} and (15) in \cite{Olver2}, \eref{f_eq} and (17) in
\cite{Olver2} are respectively identical up to scalings of
variables. The system \eref{matrix_eq} was surveyed in a previous
paper from the viewpoint of the inverse scattering method \cite{Tsuchida5}. 
The system \eref{a_eq} is a well-known matrix generalization of the 
DNLS equation of Kaup-Newell type \cite{Fordy2}. For systems 
\eref{b_eq}, \eref{c_eq}, \eref{d_eq}, \eref{e_eq} and \eref{f_eq}, 
or (12), (14)--(17) in \cite{Olver2}, we have obtained the Lax
representations by virtue of a non-commutative version of gauge 
transformations for the first time in this paper.

The systems \eref{matrix_eq}, \eref{e_eq} and \eref{f_eq} are 
interpreted as matrix generalizations of \eref{K_eq} with $\a=1$, 
$\de = 0$. The systems \eref{a_eq} and \eref{d_eq} reduce to
\eref{K_eq} with $\a =1$, $\de = -1/4$ in the commutative case. The
systems \eref{b_eq} and \eref{c_eq} correspond to \eref{K_eq} with 
$\a = 1$, $\de = 1/4$ when $\qa$ and $\ra$ are scalar-valued. The 
symmetry approach shows that matrix generalizations of \eref{K_eq} are 
essentially exhausted by the above statement up to scalings and the 
transposition \cite{Olver2}. It is remarkable that the system
\eref{K_eq} has matrix generalizations only for some choices of 
$\a$ and $\de$. This feature of matrix generalizations may be explained
from our approach in the following way. In the 
case of scalar variables, equation \eref{K_eq} is generated by the 
gauge transformation
\[
\eql{
\qsa = \frac{1}{\sqrt{\a}} \QS1 \exp \Bigl\{-2\i \frac{\de}{\a} \int^x \RS1 \QS1 \d x'\Bigr\}
\\
\rsa = \frac{1}{\sqrt{\a}} \RS1 \exp \Bigl\{2\i \frac{\de}{\a} \int^x \QS1 \RS1 \d x'\Bigr\}
}
\]
for the DNLS equation by Chen, Lee and Liu \cite{CLL}
\[
\eql{
 \i \QS1_t + \QS1_{xx} -\i \QS1 \RS1 \QS1_x = 0
\\
\i \RS1_t - \RS1_{xx} -\i \RS1 \QS1 \RS1_x = 0.
}
\]
Why cannot we generalize this transformation to the matrix case for 
the continuous parameters $\a$ and $\de \,$? As an illustrative example, 
we choose $F$ and $G^{-1}$ to satisfy
\[
F_x = \i \gamma \R1 \Q1 F \hspace{5mm} \lim_{x \to x_0} F = I
\hspace{5mm} G^{-1} = I
\]
where $\gamma$ is a scalar parameter. The explicit form of $F$ is
given by
\bea
F &=& {\cal E} \exp \Bigl\{ \i \gamma \int_{x_0}^x \R1 \Q1 \d x'\Bigr\}
\nn \\
&=& I + \sum_{n=1}^\infty (\i \gamma)^n A_n.
\nn 
\eea
Here ${\cal E}$ is the path ordering operator and
\[
A_n = \int_{x_0}^x \d x_1 \int_{x_0}^{x_1} \d x_2 \cdots 
   \int_{x_0}^{x_{n-1}} \d x_n \R1 (x_1) \Q1 (x_1) \cdots \R1 (x_n)
   \Q1 (x_n) .
\]
We can calculate the time derivative of $F$, $F_t$, with 
the help of \eref{matrix_eq}. However, it is observed that $F_t F^{-1}$
cannot be expressed in a closed form in general, i.e. it includes
infinitely multiple integrals. An exception is the case of $\gamma =
1/2$, which gives
\[
F_t = \Bigl\{ \hf (\R1_x \Q1- \R1\Q1_x ) 
+ \i \frac{1}{4}\R1\Q1\R1\Q1 \Bigr\} F
\]
and, as a result, example (d). The example explains why the
transformation \eref{tra1} is effective for finite choices of 
$F$ and $G$ in the matrix case. 

It is a well-known fact that the spatial part of the Lax formulation 
for the Chen-Lee-Liu-type DNLS equation is common with 
that for the massive Thirring model, i.e. they belong to the same 
hierarchy. Thus, employing an appropriate time part in the 
Lax formulations in correspondence with the Lax matrices $U$ 
(or $U'$) in this section, we can obtain new matrix 
generalizations 
of the massive Thirring model (see appendix). 

\section{`C-integrable' systems}
\label{C-int}

In the previous section, we have verified that all but two systems
proposed by Olver and Sokolov are `S-integrable', i.e. they have Lax 
representations and can be linearized by the inverse scattering method. In
this section, we show that two systems left for further analysis are 
`C-integrable', i.e. they can be linearized by a certain type of
 transformation of dependent variables which resembles that in the 
previous section. 

The two systems, (11) and (13) in \cite{Olver2}, are matrix
generalizations of \eref{Csys} with $\a =0$, $\beta =1$. We 
briefly summarize a solution of \eref{Csys} for investigating (11) and 
(13). The pair of equations \eref{Csys} is rewritten in a linearized
form as
\[
\eql{
\bigl( p \e^{ \a \int^x_{x_0} sp \d x'} \bigr)_t - 
\bigl( p \e^{ \a \int^x_{x_0} sp \d x'} \bigr)_{xx} = 0 
\\
\bigl( s \e^{ -\beta \int^x_{x_0} ps \d x'} \bigr)_t +
\bigl( s \e^{ -\beta \int^x_{x_0} ps \d x'} \bigr)_{xx} = 0 
}
\]
under the boundary conditions: $\lim_{x \to x_0} p = \lim_{x \to x_0} s = 0$. 
In terms of functions $y(x,t)$ and $z(x,t)$, which satisfy a pair of heat
equations
\[
 y_t - y_{xx} = 0 \hspace{5mm} z_t + z_{xx} =0
\]
and the boundary conditions, $\lim_{x \to x_0} y = \lim_{x \to x_0} z =
0$, the general solution of \eref{Csys} is given by
\[
\eql{
p = y \e^{ - \a \int^x_{x_0} sp \d x'} 
  = y \Bigl\{1+(\a - \beta) \int^x_{x_0} zy \d x' \Bigr\}^{-\frac{\a}{\a-\beta}}
\\
s = z \e^{ \beta \int^x_{x_0} ps \d x'} 
  = z \Bigl\{ 1+(\a - \beta) \int^x_{x_0} yz \d x' \Bigr\}^{\frac{\beta}{\a-\beta}}
}
\]
for $\a \neq \be$ (cf. \cite{Calo2} for $\a = -\be$) and 
\[
\eql{
p = y \e^{ - \a \int^x_{x_0} sp \d x'} 
  = y \e^{ - \a \int^x_{x_0} zy \d x'} 
\\
s = z \e^{ \a \int^x_{x_0} ps \d x'} 
  = z \e^{ \a \int^x_{x_0} yz \d x'} 
}
\]
for $\a = \be$. 

We proceed to solve (11) and (13) in \cite{Olver2} by generalizing 
the above method to the matrix case. We write (11) in
\cite{Olver2}:
\beq
\eql{
\P1_t = \P1_{xx}
\\
\S1_t = -\S1_{xx} + 2 \S1 \P1_x \S1 + 2\S1 \P1 \S1_x .
}
\label{pa_1}
\eeq
Here $\P1$ is an $n_1 \times n_2$ matrix and $\S1$ is an $n_2 \times n_1$
matrix. The boundary conditions
\[
\lim_{x \to x_0} \P1 =O \hspace{5mm}\lim_{x \to x_0} \S1 =O 
\]
are assumed for convenience. In terms of $A$ defined by
\beq
A_x = A (-\S1 \P1) \hspace{5mm} A_t = A ( \S1_x\P1 -\S1\P1_x -
\S1\P1\S1\P1) 
\label{A_eq}
\eeq
the time evolution equation for $\S1$ in \eref{pa_1} is rewritten in a 
linearized form as
\[
(A\S1)_t + (A\S1)_{xx} = O .
\]
Here the consistency condition, $A_{xt} = A_{tx}$, is checked by a
direct calculation by use of \eref{pa_1}. We introduce an 
$n_1 \times n_2$ matrix $Y(x,t)$ and an $n_2 \times n_1$ matrix
$Z(x,t)$ which satisfy a pair of matrix heat equations
\beq
Y_t -Y_{xx} = O \hspace{5mm} Z_t + Z_{xx} =O
\label{YZ}
\eeq
and the boundary conditions
\[
\lim_{x \to x_0} Y =O \hspace{5mm}\lim_{x \to x_0} Z =O .
\]
The general solution of \eref{YZ} is obtained by means of the 
Fourier transformation. Thus, if we set
\[
\P1 = Y \hspace{5mm} \S1 = A^{-1} Z
\]
this gives the general solution of \eref{pa_1}. Due to the relation 
$A_x = -Z Y$, we obtain
\[
A = I_2 - \int_{x_0}^x Z(x',t) Y(x',t) \d x'.
\]
Here we have assumed the boundary condition, $\lim_{x \to x_0} A =
I_2$, with $I_2$ being the $n_2 \times n_2$ identity matrix. 
In conclusion, an 
explicit form of the general solution of \eref{pa_1} is given by
\beq
\fl
\P1(x,t) = Y(x,t) \hspace{5mm} \S1(x,t) 
 =  \Bigl\{I_2 - \int_{x_0}^x Z(x',t) Y(x',t) 
 \d x' \Bigr\}^{-1} Z(x,t).
\label{}
\eeq

Finally, we shall derive the general solution of the only system left 
to solve, (13) in \cite{Olver2}. For this purpose, we set $n_1=n_2$ and
perform a change of the dependent variables:
\[
\P1 = A \pa  \hspace{5mm} \S1 = \sa A^{-1} .
\]
Noting the fact that \eref{A_eq} is rewritten in terms of the new
variables $\pa$ and $\sa$ as
\[
A_x = A (-\sa \pa) \hspace{5mm} A_t = A (\sa_x \pa - \sa \pa_x 
-\sa \pa \sa \pa + 2 \sa^2 \pa^2)
\]
%
%\beq
%\eql{
%(A \pa)_t - (A \pa)_{xx} =O
%\\
%(\sa A^{-1} )_t = - (\sa A^{-1})_{xx} + 2 (\sa A^{-1}) (A\pa)(\sa
%A^{-1})_x + 2 (\sa A^{-1}) (A \pa)_x (\sa A^{-1})
%}
%\label{}
%\eeq
%
we obtain the evolution equations for $\pa$ and $\sa$:
\beq
\eql{
\fl \pa_t = \pa_{xx} - 2 \sa_x \pa^2 -2 \sa \pa \pa_x 
+ 2 \sa \pa \sa \pa^2 - 2 \sa^2 \pa^3
\\
\fl \sa_t = - \sa_{xx}-2 \sa^2 \pa_x-2 \sa_x \sa \pa 
 + 2 \sa \pa \sa_x + 2 \sa \pa_x \sa 
\\ 
\fl \hspace{9.3mm}
+2 \sa \pa \sa^2 \pa + 2\sa^3 \pa^2 -2 \sa^2 \pa \sa \pa 
-2 \sa^2 \pa^2 \sa.
}
\label{pa_2}
\eeq
This is nothing but the system (13) in \cite{Olver2} up to scalings 
of variables. Thus, by virtue of the derivation in the above, the 
general solution of \eref{pa_2}, which is an alternative expression of 
(13), is obtained straightforwardly:
\beq
\eql{
\fl \pa (x,t) = A^{-1} \P1 = \Bigl\{ I - \int_{x_0}^x Z(x',t) Y(x',t) 
 \d x' \Bigr\}^{-1} Y(x,t)
\\
\fl \sa (x,t) = \S1 A = \Bigl\{ I - \int_{x_0}^x Z(x',t) Y(x',t) 
 \d x' \Bigr\}^{-1} Z(x,t) \Bigl\{ I - \int_{x_0}^x Z(x',t) Y(x',t)
 \d x' \Bigr\}.
}
\label{}
\eeq
It should be noted that all of the matrices in the above expression
are square matrices.

\section{Concluding remarks}

In this paper, we have studied matrix-valued systems of the derivative 
nonlinear Schr\"{o}dinger (DNLS) type. Applying a kind of gauge
transformation to a matrix version of the DNLS equation of the 
Chen-Lee-Liu type with a Lax pair \cite{Tsuchida5}, we have derived 
Lax representations for all but two systems proposed in
\cite{Olver2}. Hence, these systems can be linearized through the inverse 
scattering formulation and proved to be `S-integrable' in the
terminology of Calogero. As has been clarified in section 2, these
systems each are connected with the others through transformations of
the dependent variables. However, it is noteworthy that the
transformations cannot be written in a closed form in terms of 
the matrix-valued dependent variables. More explicitly, if $F$ (or $G^{-1}$) 
is not the identity, we may not express $F$ (or $G^{-1}$) for the 
examples (a)--(f) in section 2 without using the path ordering 
operator, infinitely multiple integrals, etc. 

For the two systems in \cite{Olver2} left to prove their complete 
integrability, we have shown that both of them are linearizable and
connected with each other by a change of dependent variables. The
transformations which linearize the two systems can be explicitly
written in a closed form in terms of the variables $Y$ and $Z$. This
enables us to obtain the general solutions of the two systems, which 
directly proves their `C-integrability'. 

To conclude, we have proved for the first time that 
all the matrix-valued systems proposed 
in \cite{Olver2} can be integrated by the inverse scattering method or the 
transformations of the dependent variables. The 
dependent variables of the systems take values in either square 
matrices or, more generally, rectangular matrices of arbitrary size. 
However, it is noteworthy that not all of integrable multi-component 
systems can be expressed in terms of matrix variables of arbitrary size. For 
instance, we consider the following Lax pair,
\bea
\fl U =
\i \z^2 \left[
\begin{array}{cc}
 -I_1  &  \\
    &  I_2 \\
\end{array}
\right]
+ 
\z \left[
\begin{array}{cc}
   &  \qa \\
 \ra  &   \\
\end{array}
\right]
+ \i
\left[
\begin{array}{cc}
 \hf \qa \ra &  \\
   & \hf \ra \qa \\
\end{array}
\right]
\label{U_KN}
\\
\fl V =
\i \z^4 
\left[
\begin{array}{cc}
 -2I_1 &  \\
   & 2I_2  \\
\end{array}
\right]
+\z^3
\left[
\begin{array}{cc}
  & 2\qa \\
 2\ra &  \\
\end{array}
\right]
+\i \z^2
\left[
\begin{array}{cc}
-\qa \ra  &  \\
  & \ra \qa \\
\end{array}
\right]
\nn \\
\fl  \hspace{7.5mm} +\z
\left[
\begin{array}{cc}
  & \i \qa_x + \qa \ra \qa \\
 -\i \ra_x + \ra \qa \ra &  \\
\end{array}
\right]
\nn \\
\fl \hspace{7.5mm}
+\i
\left[
\begin{array}{cc}
 \frac{\i}{2}(\qa_x \ra - \qa \ra_x) + \frac{3}{4}\qa \ra \qa \ra &  \\
 & \frac{\i}{2}(\ra \qa_x - \ra_x \qa) + \frac{3}{4}\ra \qa \ra \qa \\
\end{array}
\right] .
%\nn 
\label{V_KN}
\eea
Substituting \eref{U_KN} and \eref{V_KN} into \eref{Lax_eq}, we 
obtain the evolution equations for $\qa$ and $\ra$, 
\beq
\eql{
\i \qa_t + \qa_{xx} - \i \qa\ra_x \qa -2 \i \qa\ra\qa_x =O
\\
\i \ra_t - \ra_{xx} - \i \ra\qa_x \ra -2 \i \ra_x \qa\ra =O
}
\label{newKN}
\eeq
and the commutation relation,
\beq
[ \qa \ra, \, \qa_x \ra - \qa \ra_x] = O.
\label{commu}
\eeq
Setting
\[
\eql{
\qa = (q_1, q_2, \cdots, q_m) = \vecvar{q}
\\
\ra = {}^t(r_1, r_2, \cdots, r_m) = {}^t\vecvar{r}
}
\]
which automatically satisfies the constraint \eref{commu}, we obtain 
a system of the Kaup-Newell-type DNLS equations \cite{Tsuchida5},
\beq
\eql{
\i \vecvar{q}_t + \vecvar{q}_{xx}-\i <\vecvar{q}, \, \vecvar{r}_x>
\vecvar{q} - 2\i <\vecvar{q}, \, \vecvar{r}> \vecvar{q}_x = \vecvar{0}
\\
\i \vecvar{r}_t - \vecvar{r}_{xx}-\i <\vecvar{r}, \, \vecvar{q}_x>
\vecvar{r} - 2\i <\vecvar{r}, \, \vecvar{q}> \vecvar{r}_x = \vecvar{0}.
}
\label{vecKN}
\eeq
Here $< \; , \;>$ denotes the inner product between vectors. 
The vector Kaup-Newell system \eref{vecKN} is `S-integrable'
because it possesses the Lax pair \eref{U_KN} and \eref{V_KN} with 
changing $\qa \to \vecvar{q}$, $\ra \to {}^t\vecvar{r}$. 
However, the matrix-valued extension of \eref{K_eq} with $\a =1$, $\de=-1/4$, 
\eref{newKN}, is not `S-integrable' in general for $\qa$ and $\ra$ 
of arbitrary size. 

\ack
One of the authors (TT) 
appreciates a Research Fellowship of the Japan Society for 
the Promotion of Science for Young Scientists.

\appendix
\section*{Appendix. Massive Thirring model}
\setcounter{section}{1}
In this appendix, we show a list of matrix generalizations of the
massive Thirring model, which are respectively a member of 
the DNLS-type hierarchies studied in section \ref{S-int}. 
For this purpose, we consider Lax pairs with the following dependence
on the spectral parameter $\z$,
\bea
%\fl 
U =
\i \z^2 \left[
\begin{array}{cc}
 -I_1  &  \\
    &  I_2 \\
\end{array}
\right]
+ 
\z \left[
\begin{array}{cc}
   &  \qa \\
 \ra  &   \\
\end{array}
\right]
+ \i
\left[
\begin{array}{cc}
  U_{11} &  \\
    & U_{22} \\
\end{array}
\right]
\label{}
\eea
\bea
%\fl 
V =
\i \frac{m^2}{4 \z^2} \left[
\begin{array}{cc}
 - I_1  &  \\
   &  I_2 \\
\end{array}
\right]
+ 
\frac{m}{2\z} \left[
\begin{array}{cc}
   &  \ja \\
 \ta  &   \\
\end{array}
\right]
+ 
\i
\left[
\begin{array}{cc}
  V_{11} &  \\
    & V_{22} \\
\end{array}
\right].
\label{}
\eea
Here $U_{11}$, $V_{11}$ and $U_{22}$, $V_{22}$ are $\z$-independent 
square matrices whose size are respectively $n_1 \times n_1$ and 
$n_2 \times n_2$. $\qa$ and $\ja$ are $n_1 \times n_2$ matrices. 
$\ra$ and $\ta$ are $n_2 \times n_1$ matrices. $m$ is a nonzero
constant. 
We have derived new Lax pairs for several matrix generalizations 
of the DNLS-type equation \eref{K_eq} in section \ref{S-int}. 
To obtain matrix versions of the massive Thirring model, we have only
 to change the time part of the Lax pairs as above.
In this formulation, the new pair of potentials $\ja$ and $\ta$ appears. 
In correspondence with the choices of $U_{jj} \; (j=1,2)$, we can 
determine $V_{jj} \; (j=1,2)$ so that the compatibility condition 
\eref{Lax_eq} yields a consistent set of evolution equations. 

We can obtain four matrix generalizations of the massive Thirring 
model by the above-mentioned method. The result of the choices of 
$U_{jj}$, $V_{jj}$ $(j=1,2)$ and the corresponding evolution equations 
is listed as follows. 
\[
\fl {\rm ({\frak a})}\hspace{20mm} 
U_{11} = O \hspace{5mm} U_{22}= \hf RQ 
\hspace{5mm} V_{11} = - \hf \ja \ta \hspace{5mm} V_{22}=O
\label{}
\]
\beq
\eql{
\qa_t - \i m \ja + \i \hf \ja \ta \qa = O
\\
\ra_t + \i m \ta - \i \hf \ra \ja \ta = O
\\
\ja_x - \i m \qa + \i \hf \ja \ra \qa = O
\\
\ta_x + \i m \ra - \i \hf \ra \qa \ta = O.
}
\label{fourwave}
\eeq
\[
\fl {\rm ({\frak b})}\hspace{20mm} 
U_{11} = O \hspace{5mm} U_{22}= O
\hspace{5mm} V_{11} = - \hf \ja \ta \hspace{5mm} V_{22}= \hf \ta \ja
\label{}
\]
\beq
\eql{
\qa_t - \i m \ja + \i \hf (\qa \ta \ja + \ja \ta \qa) = O
\\
\ra_t + \i m \ta - \i \hf (\ta \ja \ra + \ra \ja \ta) = O
\\
\ja_x - \i m \qa = O
\\
\ta_x + \i m \ra = O.
}
\label{}
\eeq
\[
\fl {\rm ({\frak c})}\hspace{20mm} 
U_{11} = O \hspace{5mm} U_{22}= \hf (\ra \qa + \qa \ra)
\hspace{5mm} V_{11} = - \hf \ja \ta \hspace{5mm} V_{22}= - \hf \ja \ta
\label{}
\]
\beq
\eql{
\qa_t - \i m \ja + \i \hf (\ja \ta \qa - \qa \ja \ta) = O
\\
\ra_t + \i m \ta - \i \hf (\ra \ja \ta - \ja \ta \ra) = O
\\
\ja_x - \i m \qa + \i \hf \ja (\ra \qa + \qa \ra) = O
\\
\ta_x + \i m \ra - \i \hf (\ra \qa + \qa \ra) \ta = O.
}
\label{}
\eeq
\[
\fl {\rm ({\frak d})}\hspace{20mm} 
U_{11} = -\hf \ra \qa \hspace{5mm} U_{22}= O
\hspace{5mm} V_{11} = \hf (\ta \ja -\ja \ta) \hspace{5mm} V_{22}= \hf \ta \ja
\label{}
\]
\beq
\eql{
\qa_t - \i m \ja + \i \hf (\qa \ta \ja + \ja \ta \qa - \ta \ja \qa) = O
\\
\ra_t + \i m \ta - \i \hf (\ta \ja \ra + \ra \ja \ta - \ra \ta \ja) = O
\\
\ja_x - \i m \qa + \i \hf \ra \qa \ja = O
\\
\ta_x + \i m \ra - \i \hf \ta \ra \qa = O.
}
\label{}
\eeq
In view of the Lax matrix $U$ 
the cases (${\frak a}$)--(${\frak d}$) respectively 
correspond to \eref{matrix_eq}, 
(a), (c) and (e) in section \ref{S-int}. 
Other choices of $U$ obtained in section 
\ref{S-int} lead to systems which coincide with one of 
(${\frak a}$)--(${\frak d}$) up to the exchange of $t$ and $x$, etc. 
All of the systems (${\frak a}$)--(${\frak d}$) 
are shown to be connected with the others through the 
gauge transformations utilized in section \ref{S-int}.

The system \eref{fourwave} has been obtained 
by the authors in \cite{Tsuchida0}. The others 
seem to be new integrable systems. For the existence of products of the 
matrices in the evolution equations, $\qa$, $\ra$, $\ja$ and $\ta$ 
must be square matrices of the same size in (${\frak c}$) and 
(${\frak d}$).

\end{document}